**Full and Fractional Counting in Bibliometric Networks**

Loet Leydesdorff
* corresponding author; University of Amsterdam, Amsterdam School of Communication Research (ASCoR), PO Box 15793, 1001 NG Amsterdam, The Netherlands; email: loet@leydesdorff.net

Han Woo Park
Department of Media & Communication, Interdisciplinary Program of East Asian Cultural Studies, Interdisciplinary Program of Digital Convergence Business, Yeungnam University, 214-1 Dae-dong, Gyeongsan-si, Gyeongsangbuk-do 712-749, South Korea; email: hanpark@ynu.ac.kr

In their study entitled "Constructing bibliometric networks: A comparison between full and fractional counting," Perianes-Rodriguez, Waltman, & van Eck (2016; henceforth abbreviated as PWvE) provide arguments for the use of fractional counting at the network level as different from the level of publications. Whereas fractional counting in the latter case divides the credit among co-authors (countries, institutions, etc.), fractional counting at the network level can normalize the relative weights of links and thereby clarify the structures in the network. PWvE, however, propose a counting scheme for fractional counting that is one among other possible ones. Alternative schemes proposed by Batagelj and Cerinšek (2013) and Park, Yoon, & Leydesdorff (2016; henceforth abbreviated as PYL) are discussed in an appendix. However, our approach is not correctly identified as identical to their Equation A3. Here below, we distinguish three approaches analytically; routines for applying these approaches to bibliometric data are also provided.

As is common in social-network analysis (SNA), the co-occurrence matrix is defined by PWvE as the multiplication of the occurrence matrix by its transposed (Eq. 3: $\mathbf{U} = \mathbf{A}\mathbf{A}^\mathbf{T}$). Using an equation derived by Newman (2001), the authors posit that "(t)he number of fractional counting co-authorship links between researcher *i* and *j*, denoted by $u^*_{ij}$, is given by:"

$$u^*_{ij} = \sum_{k=1}^{N} \frac{a_{ik} a_{jk}}{n_k - 1} \qquad (1)$$

Whereas the denominator is equal to one in the case of full counting, this denominator normalizes in Eq. 1. The normalization is at the paper level ($k$): $n_k$ is the number of co-authorships of paper $k$; $N$ is the number of publications in the set. The ($n - 1$) rather than $n$ in the denominator corrects for the self-link: each author has only ($n - 1$) co-authors (Newman, 2001, p. 016132-5).

The argument is elaborated by the authors with both model and empirical examples. We focus here on the first example of a co-authorship matrix (Tables 2 and 3; their Table 3) which is based on the assumed authorship matrix in Table 1 (their Table 2, at p. 1182):

**Table 1: Authorship matrix**

|    | P1 | P2 | P3 | Total |
|----|----|----|----|-------|
| R1 | 1  | 1  | 0  | 2     |
| R2 | 1  | 0  | 1  | 2     |
| R3 | 1  | 1  | 0  | 2     |
| R4 | 0  | 0  | 1  | 1     |
| Total | 3 | 2 | 2 |       |

**Table 2: Full counting**

|    | R1 | R2 | R3 | R4 | Tot |
|----|----|----|----|----|-----|
| R1 |    | 1  | 2  | 0  | 3   |
| R2 | 1  |    | 1  | 1  | 2   |
| R3 | 2  | 1  |    | 0  | 3   |
| R4 | 0  | 1  | 0  |    | 1   |
| To | 3  | 2  | 3  | 1  | 9   |

**Table 3: Fractional counting PWvE**

|    | R1  | R2  | R3  | R4  | Total |
|----|-----|-----|-----|-----|-------|
| R1 |     | 0.5 | 1.5 | 0.0 | 2.0   |
| R2 | 0.5 |     | 0.5 | 1.0 | 2.0   |
| R3 | 1.5 | 0.5 |     | 0.0 | 2.0   |
| R4 | 0.0 | 1.0 | 0.0 |     | 1.0   |
| To | 2.0 | 2.0 | 2.0 | 1.0 | 7     |

**Our alternative approach**

The first document P1 is co-authored by three authors, each of whom would receive one-third point of the credit when counted fractionally (Table 4). In the fractionated co-authorship matrix, the cell value {R1, R2} is accordingly 1/3 * 1/3 = 1/9 or 0.11 (Table 5).

**Table 4: Fractional counting**

|    | P1   | P2   | P3   | Total |
|----|------|------|------|-------|
| R1 | 0.33 | 0.50 | 0.00 | 0.83  |

**Table 5: Fractionally counted co-authorship matrix**

|    | R1   | R2   | R3   | R4   | Total |
|----|------|------|------|------|-------|
| R1 | 0.36 | 0.11 | 0.36 | 0.00 | 0.83  |

| | | | | | | | | | | |
|---|---|---|---|---|---|---|---|---|---|---|
| R2 | 0.33 | 0.00 | 0.50 | 0.83 | R2 | 0.11 | 0.36 | 0.11 | 0.25 | 0.83 |
| R3 | 0.33 | 0.50 | 0.00 | 0.83 | R3 | 0.36 | 0.11 | 0.36 | 0.00 | 0.83 |
| R4 | 0.00 | 0.00 | 0.50 | 0.50 | R4 | 0.00 | 0.25 | 0.00 | 0.25 | 0.50 |
| Total | 1 | 1 | 1 | 3 | Total | 0.83 | 0.83 | 0.83 | 0.50 | 2.98 |

Eq. 2 formalizes this approach, as follows:

$$u*_{ij} = \sum_{k=1}^{N} \frac{a_{ik}a_{jk}}{n_{ik}n_{jk}} = \sum_{k=1}^{N} \frac{a_{ik}a_{jk}}{n_k^2} \quad (2)$$

Note that our values are smaller than those of PWvE because the value of the denominator ($n^2$) is larger than ($n - 1$). Whereas PWvE count a total of seven for three papers, we count three (after rounding). In other words, this method is consistent. Table 5 is also provided as Table A1 in the Appendix of PWvE (at p. 1194), but without the diagonal values so that this consistency is not noticed.

**Directed versus undirected networks**

In both methods, the numerator ($a_{ik}a_{jk}$) for paper $k$ and authors $i$ and $j$ is based on the assumption that the relation of $i$ with $j$ is counted as one arrow in addition to the reverse relation of $j$ with $i$. While relations are counted bi-directionally in SNA, from a bibliometrics perspective, co-authorship is conceptualized as a single edge instead of opposing arcs. When we accept the argument of PWvE to correct for the self-relation, only one of the two arcs is being corrected. The denominator would then be $n * (n - 1)$ instead of $n^2$. The resulting matrix (not shown here; but see Table 6 below) contains somewhat higher values than the ones in Table 5 and sums to 3.35.

The distinction of bilateral arcs in SNA versus undirected networks in bibliometrics raises the question of how one corrects for the double values of arrows in both directions. In the co-occurrence matrix, this doubling is represented by the equivalency between the upper and lower triangles. Since these triangles contain $n * (n – 1) / 2$ cells, we recommend using this value for the denominator (if consistency is not a major issue given the research question)., Note that the multiplication by two does makes no difference for the structure of the matrix, since it applies equally to all cells. Table 6 provides the resulting matrix; Eq. 3 the formalization (Cerinšek & Batagelj, 2015, at p. 987). As expected, the values are twice as large.

Table 6: Co-authorship network fractionally counted using $n * (n – 1) / 2$

|      | R1   | R2   | R3   | R4   | TOTAL |
|------|------|------|------|------|-------|
| R1   | 0.86 | 0.26 | 0.86 | 0.00 | 1.98  |
| R2   | 0.27 | 0.90 | 0.27 | 0.63 | 2.07  |
| R3   | 0.86 | 0.26 | 0.86 | 0.00 | 1.98  |
| R4   | 0.00 | 1.00 | 0.00 | 1.00 | 2.00  |
| Total| 1.86 | 1.86 | 1.86 | 1.13 | 6.71  |

$$u^*_{ij} = \sum_{k=1}^{N} \frac{a_{ik} a_{jk} * 2}{n_{jk} * (n_{ik} - 1)} = \sum_{k=1}^{N} \frac{a_{ik} a_{jk} * 2}{n_k * (n_k - 1)} \tag{3}$$

**Discussion and conclusion**

The algorithm for fractional counting can be used at different levels of aggregation, such as networks of authors, institutions, countries, etc. Whereas an author-document matrix is binary at the level of authors, at other levels of aggregation the matrices can be valued. Using Eq. 1 in that case, one divides a squared number in the numerator by a linear function in the denominator. In the binary case, this is no problem since $1^2 = 1$ and $0^2 = 0$. In the non-binary case, however, this difference in the dimensionality may lead to unintended effects. Using Eqs. 2 or 3, one corrects for this potential imbalance. Only Eq. 2 is consistent, but this equation has the conceptual disadvantage of including the self-relation.

We have implemented the three equations in a routine available at http://www.leydesdorff.net/software/fractionate/index.htm for data downloaded from the Web of Science. (Using Eq. 3 is the default.) The user is first prompted for the level of aggregation: (a)uthor, (i)nstitution or (c)ountry. The resulting file mtrx.net contains the whole-number-counted co-occurrence matrix and the file fmtrx1.net, fmtrx2.txt, or fmtrx3.txt containing the fractionally counted matrix corresponding with using Eqs. 1, 2, or 3, respectively. The files are in the Pajek edgelist format so that there are no limitations of size.

In PYL, we noted that searches at the internet or using search engines in databases do not retrieve co-occurrence numbers ($a_{ik} a_{jk}$) as discussed here, but the minimal overlaps using a Boolean AND (Morris, 2005, at p. 22; Leydesdorff & Vaughan, 2006, at pp. 1626f.). Based on this other co-occurrence data, one may need a different approach to the problem of fractional counting of the network (Zhou & Leydesdorff, 2016).

We agree with PWvE that fractionation solves a problem in appreciating co-authorship networks. The modularity is reduced; the resulting clusters tend to be larger and sorted more clearly apart. In previous research, clarification of the co-occurrence network was often induced by using the cosine or another similarity measure for the normalization (e.g., Ahlgren *et al.*, 2003; Leydesdorff, 2008; Leydesdorff & Vaughan, 2006; Schneider & Borlund, 2009; Waltman & van

Eck, 2013). In our opinion, fractional counting solves this problem more elegantly. The empirical elaboration, however, remains a subject for further research.

**Acknowledgement**
We are grateful to Wouter de Nooy for comments on a previous draft.**Acknowledgement**
We are grateful to Wouter de Nooy for comments on a previous draft.**Acknowledgement**
We are grateful to Wouter de Nooy for comments on a previous draft.